# Resource Allocation Algorithm for V2X communications based on SCMA

Wei wu, Linglin Kong and Tong Xue

Communication Research Center, Harbin Institute of Technology, Harbin, 150080, China

Emails:{kevinking, tengerskong}@hit.edu.cn, xtong940213@163.com

**Abstract:** In this paper, we propose a resource allocation algorithm for V2X communications based on Sparse Code Multiple Access(SCMA). By analyzing the interference model in the V2X scenario, we formulate the problem which deals with resource allocation to maximize the system throughput. A graph color-based user cluster algorithm combined with resource allocation algorithm based on both result of clustering and SINR is presented to solve the problem. The simulation results indicate that the throughput performance of system based on SCMA is superior to which based on OFDMA, and the proposed algorithm can improve the system throughput and the number of access users.

## I. Introduction

With the increasing of the number of vehicles and the development of the Internet, the research of Internet of Vehicles(IOV) becomes a research focus [1]. Low latency and high reliability are critical requirements for the services, especially safety applications, in the IOV [2]. However, in the scenario of IOV, the vehicles are densely distributed, fast moving and the number of users is large. It is difficult for the base station to obtain full channel state information(CSI). It will lead to a longer resource scheduling time, unreliable channel information.

V2X is the general term for Vehicle-to-Vehicle(V2V), Vehicle-to-Infrastructure (V2I) and vehicle-to-device(V2D). The current solution for V2X is based on the IEEE 802.11p standard ad-hoc communications and back-end communications based on the Long-Term-Evolution(LTE). However, both of these solutions cannot guarantee quality of service (QoS) of IOV applications [3].

Device-to-Device (D2D) communications underlay LTE networks can be seen as one of the ways to solve the problem [4]. In D2D communication underlay LTE networks, two adjacent users (UEs) can communicate without the forward of base station(BS). D2D communications can reduce the delay and improve the reliability [5]. In V2X, D2D can be employed to achieve V2V communication. However, the performance improvement is limited when introducing D2D into V2X due to the interference problem resulted from the reusing of radio resources D2D users(D-UEs) and cellular users(C-UEs).

In [6], [7], the radio resource management (RRM) problem of the D-UEs reused the resources of C-UEs was analyzed, but only one case that one D-UE reusing the resources of one C-UE is considered. In [8], the authors proposed an RRM algorithm based on cluster-based D2D and cellular hybrid networks. This algorithm was based on orthogonal multiple access. In [9], the authors indicated that non-orthogonal multiple access technology can increase the number of connections, improve the system throughput and spectral efficiency and reduce the end-to-end delay. Therefore, V2X communication based on non-orthogonal multiple access becomes a valuable research topic in IOV.

In this paper, we consider a single cell hybrid network which includes C-UEs and V-UEs. We utilize Sparse Code Multiple Access in cellular communication to improve the number of cellular users [10]. The V-UEs reuse the resources of C-UEs. To reduce the interference between the connected V-UEs, we employ the graph coloring method to cluster the users so that several users in a cluster reuse the same resources with little interference. To maximum the sum rate of the cell, we propose a resource allocation algorithm based on SINR selection.

The rest of the paper is organized as follows. In Section II, we provide a system model, including the hybrid network and SCMA access scheme. Section III provides the formulation of problem. The graph-colored cluster selected method and resource allocation scheme are presented in Section IV. In Section V, we show the performance of our algorithm by simulation. Section is the conclusion.

## II. System Model

### A. Cell Model

The system scenario is a hybrid network consisted of C-UEs and V-UEs. C-UEs communicate with the forward of BS, i.e. V2I mode. V-UEs communicate directly, i.e. V2V mode, as shown in Figure 1. Note that $\mathcal{V} = \{1,2,\cdots,K\}$, represents the set of V-UEs, while $\mathcal{C} = \{1,2,\cdots,N\}$ represents the set of C-UEs. $K$ and $N$ represents the number of V-UEs and C-UEs respectively. Resource centrally control by BS. There is interference between C-UEs and V-UEs pairs when they reuse the same resource. In order to simplify the model, this paper assumes that both sides of the V2V communication must be in the same cell.

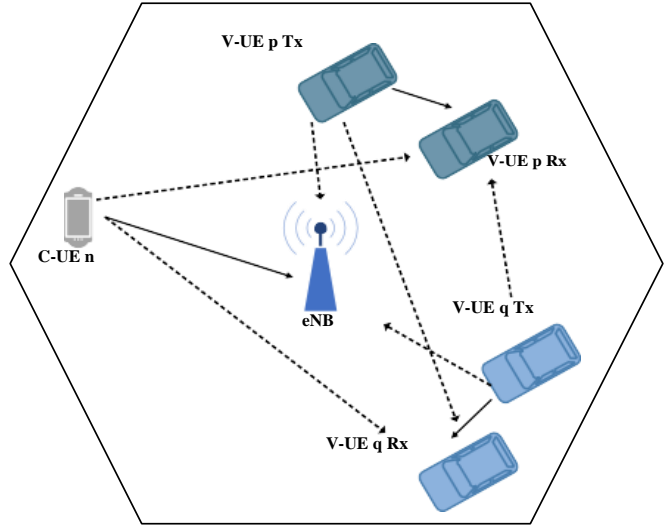

**Fig. 1** Channel sharing and interference model of V2X communications

**B. Multiple Access Scheme**

In order to increase the number of access users and the system throughput, we adopt SCMA as the multiple access scheme of C-UEs in this paper. SCMA maps the $\log_2 D$ bits to the $M$-dimensional codewords via the encoder. $D$ represented the number of codewords. The SCMA encoder of user $k$ can be expressed as:

$$x_k = f_k(\mathbf{b}) \tag{1}$$

Where $\mathbf{b}$ is the input binary bit stream, and $f_k$ is the mapping function of user $k$. $\mathbf{x}_k$ is an $M$-dimensional sparse codeword vector whose non-zero dimension $M_C < M$. The maximum number of codebooks is:

$$J = \binom{M_C}{M} \tag{2}$$

If $L$ is the number of resource blocks(RBs), the overload factor can be defined as OF = J/L. Figure 2 shows an example of SCMA encoder. The signals of 6 UEs are overlaid and spread over 4 RBs. Therefore, the overload factor is 1.5.

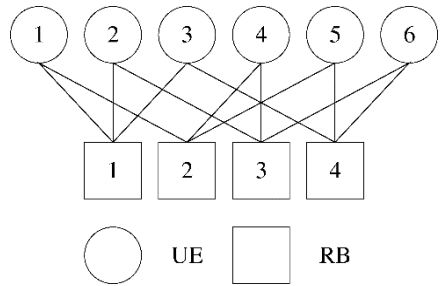

**Fig. 2** An example of SCMA encoder

Since the V-UEs which reuse the resources of the C-UEs may interfere C- UEs, the number of the V-UEs which reused the same C-UE resource should be limited. The V-UEs use orthogonal multiple access to reduce interference.

## III. Problem Formulation

In this paper, D2D communication is centrally controlled by BS. BS can obtains CSI through the measurement reports from the users. It assumes that $K$ V2V pairs reuse the same resource of $N$ C-UEs in uplinks. When the UEs adopt SCMA, as the example in Fig. 2, UE1, UE2 and UE3 use the same RB. Hence, there are co-channel interference with each other. In addition, the uplink signal of C-UE may interfere the V-UE Rx, the V-UE Tx may interfere the BS, and V2V pairs may interfere each other, when reuse occurs. Therefore, it is very necessary for BS to take measures to control the interference.

Our problem formulation deals with resource allocation for both V-UEs and C-UEs. The optimization objective is to maximize the sum rate of the cell. According to the interference analyzation and Shannon capacity, our problem can be formulated as a sum rate optimization:

$$\max\left(\sum_{i=1}^{N} log_2(1+\gamma_{CUE-i}) + \sum_{j=1}^{K} log_2(1+\gamma_{VUE-j})\right) \quad (3)$$

subject to:

$$\gamma_{CUE-i} = \frac{P_i^c h_{ci,bs}}{\sum_{j=1}^{K} P_j^v h_{bs,vj} x_{i,j} + \sum_{k=1,k\neq i}^{N} P_j^c h_{ci,cj} y_{i,j} + N_0} \geq SINR_{CUE-min}, \forall i \in C \quad (4)$$

$$\gamma_{VUE-j} = \frac{P_j^v h_{vl,vj}}{\sum_{l=1,l\neq j}^{K} P_l^v h_{vl,vj} z_{l,j} + \sum_{i=1}^{N} P_i^c h_{ci,vj} x_{i,j} + N_0} \geq SINR_{VUE-min}, \forall j \in V \quad (5)$$

$$0 \leq P_i^c \leq P_{max}^c, \forall i \in C \quad (6)$$

$$0 \leq P_j^v \leq P_{max}^v, \forall j \in V \quad (7)$$

$$0 \leq i \leq N_{RB} \cdot OF \quad (8)$$

Where $\gamma_{CUE-i}, \gamma_{VUE-j}, P_i^c$ and $P_j^v$ are the SINR and transmit power of C-UE $i$ and V-UE $j$, respectively. $x_{i,j}$, $y_{i,j}$ and $z_{l,j}$ are reuse indicator with binary values. If V-UE $j$ reuses the same RB of C-UE $i$, then $x_{i,j} = 1$, otherwise 0; if C-UE $i$ and C-UE $k$ reuse the same RB, then $y_{i,k} = 1$, otherwise 0; likely, if V-UE $l$ and V-UE $j$ reuse the same RB, then $z_{l,j} = 1$, otherwise 0. $P_{max}^c, P_{max}^v, SINR_{CUE-min}$ and $SINR_{VUE-min}$ are the maximum transmission power and minimum SINR requirement of C-UEs and V-UEs, respectively. Besides, $N_0$ is Gaussian White Noise; $N_{RB}$ represents the number of RBs in the system.

Constraints (4) and (5) ensure that the C-UE and the V-UE should meet their own quality of service (QoS) requirements. Constraints (6) and (7) indicate that the transmit power of C-UE and the V-UE should not exceed their respective maximum transmit power. Constraint (8) indicates that the number of active C-UEs accessing the base station can not greater than the product of the number of RBs in the system and the overload factor.

# IV. Graph Coloring-based Cluster Algorithm and Resource Allocation Algorithm

## A. Graph Coloring-based Cluster Algorithm

We can make multiple V-UE pairs simultaneously reuse a same C-UE resource to improve the system capacity and spectral efficiency. A graph-based clustering algorithm is adopted to achieve this target. The nodes in the interference graph denote a V-UE pair, while different colors representing different clusters. If there are interference between V-UE pair $i$ and $j$, we can't assign V-UE pair $i$ and $j$ into the same cluster, i.e., node $i$ and node $j$ can't be the same color. So we should use as few colors as possible in the interference graph to minimize the total number of clusters. To achieve this goal, our algorithm firstly colors the most interfered node, and then consider the node which is the second more interfered. According the interference relationship amoug the current node and already colored nodes to determine its color, and repeat this loop until all nodes are colored successfully.

## B. Resource Allocation Algorithm based on QoS

After the clustering is completed, we allocate resources of C-UE with best SINR C-UE to a V-UE cluster, and determine whether the SINR is high enough for Qos after reuse. If the condition is satisfied, the next V-UE cluster can continue to reuse its resource. Otherwise, the cluster chooses the next C-UE to reuse until the process ends.

**Algorithm 2.** Resource Allocation Algorithm

1: **Initialization**: $c_i \in C$: the set of C-UEs are sorted in descending order of SINR
2: $v_j \in V = \{1,2,\cdots,K\}$: the set of V-UE pairs
3: $v_{j\prime} \in V' = \{1,2,\cdots,K'\}$: the set of V-UE pairs which have been clustered
4: **if** $V \neq \emptyset$ or $c_i \neq 1$ **then**
5:    **for** $i = 1: N$ **do**
6:       **for j** $= 1 : K'$ **do**
7:          Let cluster $j'$ reuse the RB of C-UE $i$, calculate new $\gamma_{CUE-i}$ and $\gamma_{VUE-j\prime}$
8:          **if** $\gamma_{CUE-i} \geq SINR_{CUE-min}$ and $\gamma_{VUE-j\prime} \geq SINR_{VUE-min}$ **then**
9:             $j' = j' + 1$
10:         **else**
11:             $i = i + 1$
12:         **end if**
13:       **end for**
14:    **end for**
15: **end if**

**Algorithm 1.** Graph Coloring-based Cluster Algorithm
───────────────────────────────────────────────
 1: **initialization**: $index \in \{1,2,\cdots,K\}, K \geq 2$
 2: *Cluster*: the set of all clusters; create a new blank V-UE cluster $Cluster.(1)$, put it into *Cluster*
 3: **if** V-UE $j$ interferes V-UE $i$ **then**
 4:   $interf_{i,j} = 1$
 5: **else**
 6:   $interf_{i,j} = 0$
 7: **end if**
 8: **for** $i = 1:K$ **do**
 9:   $\sum_{j=1,j\neq i}^{K} interf_{i,j} = Interf_i$, put $(i, Interf_i)$ into $InterfSize$
10: **end for**
11: Sort *InterfSize* in descending order of size $Interf_i$, store in *SortedSize*
12: **for** $j = 1:K$ **do**
13:   $index = SortedSize(j).second$
14:   **for** $k = Cluster.(1) : Cluster.final$ **do**
15:     **if** $Cluster.(K) \cap Interf_{index} = \emptyset$ **then**
16:       put V-UE *index* into *Cluster.(K)*, **then** break
17:     **else**
18:       V-UE *index* is not assigned to any cluster in *Cluster*, create *Cluster.(K+1)* put V-UE *index* into this cluster and *Cluster*
19:     **end if**
20:   **end for**
21: **end for**
───────────────────────────────────────────────

## V. Simulation Results

In the simulation, we constructed a single cell. The system is a hybrid communication network with C-UEs and V-UE pairs. Two kinds of users are randomly distributed in the cell. In addition, the parameters of SCMA are set that sub-carrier number $L = 4$, non-zero elements number $Mc = 2$, codebook number $J = 6$. The other parameters are shown in Table 1.

Firstly, we compare the throughput performance of different multiple access methods when the number of RBs is same. Figure 3 shows that SCMA has a better performance than OFDMA. Since the overload feature of SCMA, it allows more users access the system. It leads to a larger average throughput.

In Figure 4, it indicates the total throughput of C-UEs is gradually reduced due to the reusing interference of the V-UE pairs, and the overall throughput of the V-UE pairs increases with the growth of number of V-UE pairs. Meanwhile, V-UE pairs bring more interference, so the total throughput of the two kinds of UEs firstly rise and then decline in the trend.

**Table 1.** Simulation Parameters

| Parameters | Value |
|---|---|
| Radius of cell | 250m |
| Carrier frequency | 2GHz |
| System bandwidth | 20MHz |
| C-UE transmit power | 20dBm |
| V-UE transmit power | 17dBm |
| Noise power spectral density | -174dBm/Hz |
| V2V link distance | 25m |

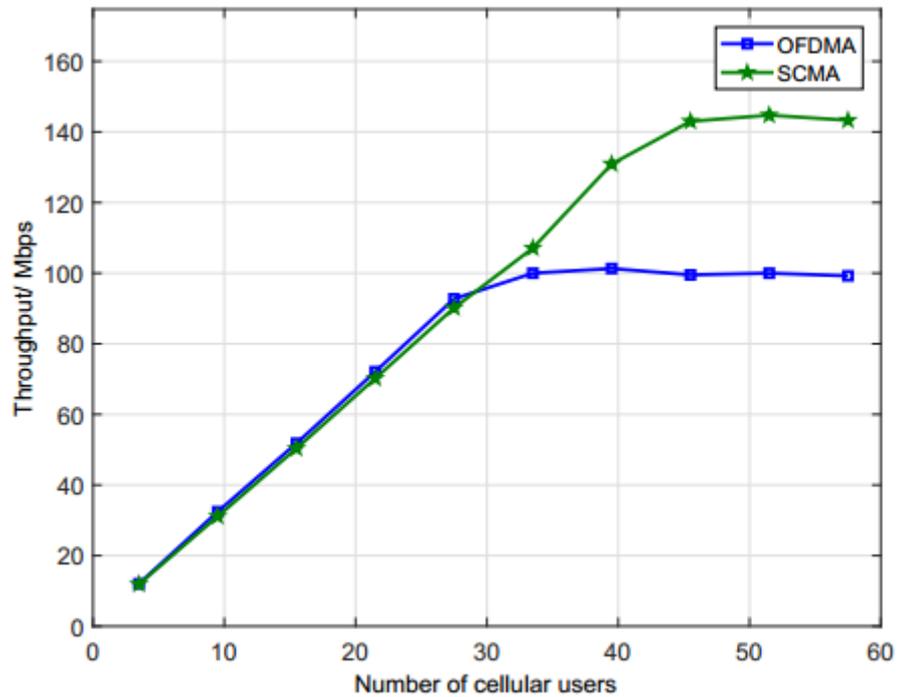

**Fig. 3** Throughput vs C-UE number with different multiple access scheme

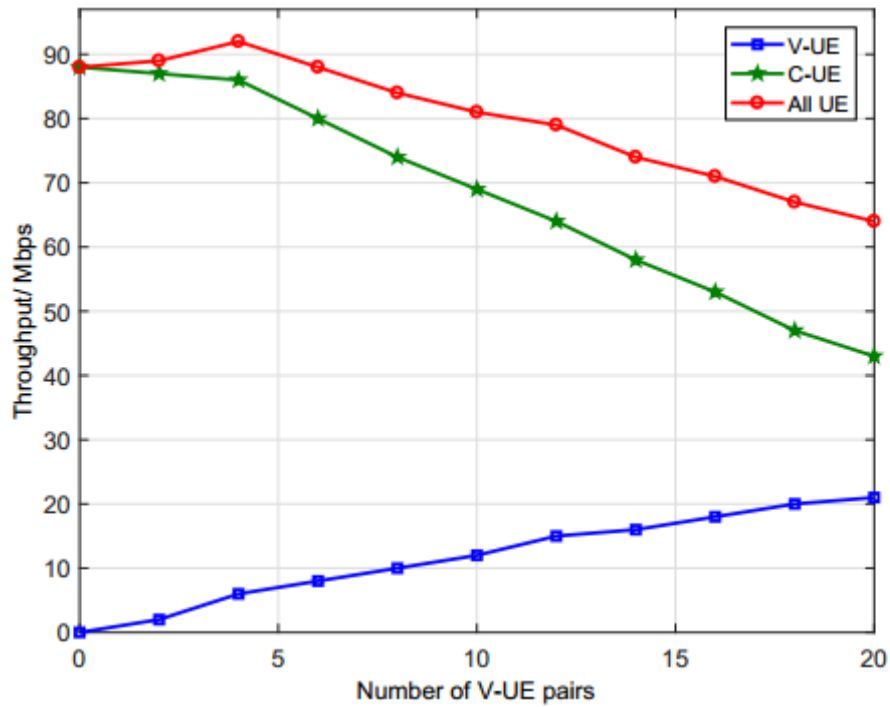

**Fig. 4.** Throughput vs V-UE pairs number

## VI. Conclusion

In this paper, we consider a hybrid network consisting of C-UEs and V-UE pairs in a single cell. To increase the number of access users, we adopt SCMA for C-UEs. The simulation results show that there is a better performance than OMA. Then, we propose a V-UEs clustering algorithm based on graph coloring theory to minimize the mutual interference of V-UEs that reuse a same RB in the same cluster. Further, we propose a resource allocation algorithm based on SINR to maximize the sum rate. The simulation results show that our algorithm can increase system throughput, as well as the number of access users.